\documentstyle[prd,aps,preprint,tighten]{revtex}
\begin{document}

\title{The theoretical significance of $G$\footnote{Talk given at the 
conference ``The Gravitational Constant: Theory and Experiment 200 
years after Cavendish'' (London, 23-24 November 1998); to appear in 
Measurement, Science and Technology, 1999.}}
\author{Thibault Damour}
\address{Institut des Hautes Etudes Scientifiques, 91440 
Bures-sur-Yvette, 
France}

\maketitle

\begin{abstract}
The quantization of gravity, and its unification with the other 
interactions, is one of the greatest challenges of theoretical physics. 
Current ideas suggest that the value of $G$ might be related to the 
other fundamental constants of physics, and that gravity might be 
richer than the standard Newton-Einstein description. This gives added 
significance to measurements of $G$ and to Cavendish-type experiments.
\end{abstract}

\vglue 10mm

Cavendish's famous experiment \cite{cavendish}, carried out in 1798 
using an apparatus conceived by the Reverend John Michell, gave the 
first accurate determination of the strength of the gravitational 
coupling. We refer to the beautiful review of Everitt \cite{cwfe} for 
an authorative discussion of this classic experiment. At the 95\% 
confidence level, the value obtained by Cavendish for the mean density 
of the Earth was $5.48 \pm 0.10$ (the modern value being 5.57) \cite{cwfe}. 
This corresponds to a fractional precision of 1.8\%. As is discussed in 
detail in the other contributions to this Cavendish bicentennial 
conference, our present knowledge of the value of Newton's constant $G$ 
seems to be uncertain at the $\sim 10^{-3}$ level. This contrasts very 
much with our knowledge of other fundamental constants of physics (e.g. 
$\alpha_{\rm e.m.} = e^2 / (4\pi \hbar c)$, and particle masses) which are 
known with a part in a million precision, or better. [One can note, 
however, that the strong coupling constant, at the $Z$-boson mass 
scale, $\alpha_s (m_Z)$ is known only with a fractional precision of 
1.7\% \cite{caso}.] The purpose of this contribution is to discuss 
briefly the significance of the value of $G$, and more generally of 
Cavendish experiments, within the current framework of theoretical 
physics.

Let us immediately note that, as any other fundamental constant of 
physics, $G$ should be measured with state-of-the-art precision, even 
if the significance of its value within the framework of physics were 
unknown (or small). But the main point I wish to make here is that many 
theoretical developments of twentieth' century physics suggest that 
there is an especially deep significance attached to the value of $G$. 
This gives, therefore, all the more importance to measurements of $G$. 
[Though, as we shall see, our current theoretical understanding is 
incomplete and cannot yet make full use of any precise value of $G$.]

As a starting point, let us remind the reader that the strength of 
gravity is strikingly smaller than that of the three other known 
interactions. Indeed, in quantum theory, the strengths of the 
electromagnetic, weak and strong interactions are measured by three 
dimensionless numbers $\alpha_1$, $\alpha_2$, $\alpha_3$ which are 
smaller but not much smaller than unity. Here, $\alpha_i \equiv g_i^2 / 
(4\pi \hbar c)$, with $i=1,2,3$, where $g_1$, $g_2$, and $g_3$ are the 
coupling constants of the gauge groups $U(1)$, $SU(2)$ and $SU(3)$, 
respectively\footnote{Here, $\alpha_1 \equiv (5/3) \alpha_Y$ with $Y$ 
being the weak hypercharge $(Y (e_R) = 1)$. The usual fine-structure 
constant $\alpha = e^2 / (4\pi \hbar c) \simeq 1/137$ corresponds to a 
combination of $\alpha_Y$ and $\alpha_2$.}. The values of the 
$\alpha_i$'s depend on the energy scale at which they are measured, 
i.e. they depend on the distance scale\footnote{We recall that in 
relativistic quantum mechanics (using $c=1$) an energy scale $E$ 
corresponds to a distance scale $L_E = \hbar / E \simeq 0.2$ fermi 
$(E/{\rm Ge} V)^{-1}$.} on which the interaction is being probed. [For 
instance, the strong coupling constant $\alpha_3$, measuring the 
strength of Quantum Chromodynamics (QCD), is of order unity at the 
energy scale $\Lambda_{\rm QCD} \sim 200 \, {\rm Me} V$, and becomes 
small at high energy scales, i.e. at very short distances.] The numerical values 
of the $\alpha_i$'s at the energy scale 
defined by the mass of the $Z$ boson, $m_Z \approx 91 \, {\rm Ge} V$, 
are \cite{polonsky}
\begin{equation}
\alpha_1 (m_Z) = \frac{1}{58.97 \pm 0.08} \, , \ \alpha_2 (m_Z) = 
\frac{1}{29.61 \pm 0.13} \, , \ \alpha_3 (m_Z) = \frac{1}{8.3 \pm 0.5} 
\, . \label{eq1}
\end{equation}
When the energy scale $\mu$ increases, $\alpha_1 (\mu)$ and $\alpha_2 (\mu)$ 
increase, while $\alpha_3 (\mu)$ decreases. It seems (if one makes extra 
assumptions about the existence and spectrum of new 
(supersymmetric) particles at higher energies) that the three gauge 
couplings unify to a common numerical value
\begin{equation}
\alpha_1 (m_U) \simeq \alpha_2 (m_U) \simeq \alpha_3 (m_U) \equiv 
\alpha_U \simeq \frac{1}{25} \label{eq2}
\end{equation}
at a very high energy scale
\begin{equation}
m_U \simeq 2 \times 10^{16} \, {\rm Ge} V \, . \label{eq3}
\end{equation}

By contrast with the numerical values (\ref{eq1}) or (\ref{eq2}), the 
corresponding ``gravitational fine-structure constant'', $\alpha_g (m) 
\equiv G \, m^2 / \hbar c$, obtained by noting that the gravitational 
interaction energy $G \, m^2 / r$ is analogous to the electric one $e^2 
/ (4\pi r)$, is strikingly small, $\alpha_g (m) \sim 10^{-40}$, when 
$m$ is taken to be a typical particle mass. Indeed,
\begin{equation}
\alpha_g (m) \equiv \frac{G \, m^2}{\hbar c} \simeq 6.707 \times 10^{-39} 
\left( \frac{m}{{\rm Ge} V} \right)^2 \, . \label{eq4}
\end{equation}

For a long time, this enormous numerical difference was viewed as a 
challenge. At face value, it seems to imply that gravity has nothing to 
do with the three other interactions. However, some authors tried to 
find a natural origin for numbers as small as (\ref{eq4}). In 
particular, Landau \cite{landau} conjectured that the very small value 
of $\alpha_g$ might be connected with the value of the fine-structure constant 
$\alpha = [137.0359895 (61)]^{-1}$ by a formula of the type $\alpha_g 
\simeq A \exp (-B / \alpha)$, with $A$ and $B$ being numbers of order 
unity. More recently, 't Hooft \cite{thooft} resurrected this idea in 
the context of instanton physics, where such exponentially small 
factors appear naturally. He suggested that the value $B = \pi / 4$ was 
natural, and he considered the case where $m=m_e$, the electron mass. 
It was noted (for fun) in Ref.~\cite{corfu} that the simple-looking 
value $A = (7\pi)^2 / 5$ happens to give an excellent agreement with 
the experimental value of $G$. Namely, if we define (for fun) a 
simple-looking ``theoretical'' value of $G$ by
\begin{equation}
\alpha_g^{\rm theory} (m_e) \equiv \frac{G^{\rm theory} \, m_e^2}{\hbar c} 
\equiv \frac{(7\pi)^2}{5} \, \exp \left( - \frac{\pi}{4\alpha} \right) \, , 
\label{eq5}
\end{equation}
one finds that it corresponds to $G^{\rm theory} = 6.6723458 \times 
10^{-8} {\rm cm}^3 \, g^{-1} \, s^{-2}$, in excellent agreement with 
the CODATA value: $G^{\rm CODATA} / G^{\rm theory} = 1.00004 \pm 
0.00013$ !

The first aim of this exercise was to exhibit one explicit example of a 
possible theoretical prediction for $G$. The second aim is to serve as 
an introduction to the currently existing ``predictions'' for the value 
of $G$ which are numerically inadequate, but which are conceptually 
important. Indeed, the main message of the present contribution is that 
the gravitational interaction is currently believed to play a central 
role in physics, and to unify with the other interactions at a very 
high energy scale. The main argument is that gravity, like the other 
interactions, should be described by quantum theory. However, 
quantizing the gravitational field has turned up to be a much more 
difficult task than quantizing the other interactions. Let us recall 
that the electromagnetic interaction was quantized in the years 
1930-1950 (QED), and that the weak and strong interactions were 
quantized in the 70's and 80's (Standard Model of weak interactions and 
QCD). The methods used to quantize the electroweak and strong 
interactions are deeply connected with the fact that the (quantum) 
coupling constants of these interactions, $\alpha_i = g_i^2 / (4\pi 
\hbar c)$, are dimensionless. By contrast, we see from Eq.~(\ref{eq4}) 
that the quantum gravitational coupling constant $G / \hbar c$ has the 
dimension of an inverse mass squared, or (using the correspondence $L_E 
= \hbar / E$) of a distance squared. This simple fact has deep 
consequences on the quantization of gravity. It means that gravity 
becomes very strong at high energies, i.e. at short distance scales. 
This is directly apparent in Eq.~(\ref{eq4}). If we consider a quantum 
process involving the mass-energy scale $\mu$, the associated 
dimensionless analog of the fine-structure constant will be $\alpha_g 
(\mu) = G \, \mu^2 / \hbar c$ and will grow quadratically with $\mu$. 
This catastrophic growth renders inefficient the (renormalizable quantum 
field theory) methods used in the quantization of the other 
interactions. It suggests that gravity defines a maximum mass scale, or 
a minimum distance. There is, at present, only one theory which, 
indeed, contains such a fundamental length scale, and which succeeds 
(at least in the perturbative sense) in making sense of quantum 
gravity: namely, String Theory. This theory (which is not yet 
constructed as a well defined, all encompassing framework) contains no 
dimensionless parameter, and only one dimensionful parameter $\alpha' = 
\ell_s^2 = m_s^{-2}$ where $\ell_s$ is a length, and $m_s$ a mass (we 
henceforth often use units where $\hbar = 1 = c$). In the simplest case 
(where the theory is perturbative, and no large dimensionless numbers 
are present), String Theory makes a conceptually very elegant 
prediction: it predicts that the ``fine-structure constants'' of {\it 
all} the interactions, including the gravitational one, must become 
equal at an energy scale of the order of the fundamental string mass 
$m_s$. In other words, it predicts (in the simplest case) that
\begin{equation}
\alpha_g (m_U) \simeq \alpha_1 (m_U) \simeq \alpha_2 (m_U) \simeq 
\alpha_3 (m_U) \ \hbox{at} \ m_U \sim m_S \, . \label{eq6}
\end{equation}
This yields $G \sim \alpha_U / m_U^2$. Taking into account some 
numerical factors \cite{kaplu} yields something like ($\gamma$ denoting 
Euler's constant)
\begin{equation}
\frac{G}{\hbar c} \simeq \frac{e^{1-\gamma}}{3^{3/2} \, 4\pi} \ 
\frac{\alpha_U}{m_U^2} \, . \label{eq7}
\end{equation}
When one inserts the ``experimental'' values (\ref{eq2}) and 
(\ref{eq3})  for $\alpha_U$ and $m_U$, one finds that the R.H.S. of 
Eq.~(\ref{eq7}) is about 100 times larger than the actual value of $G$. 
Many attempts have been made to remedy this discrepancy \cite{dienes}. 
However, the main message I wish to convey here is that modern physics 
tries to unify gravity with the other interactions and suggests the 
existence of conceptually important links, such as Eq.~(\ref{eq7}), 
between $G$ and the other coupling constants of physics. It is quite 
possible that, in the near future, there will exist a better prediction 
for $G$. 

I wish to mention that the exponential-type relations 
(\ref{eq5}) between $G$, $\alpha$ and the particle mass scales are also 
(roughly) compatible with the type of unification predicted by string 
theory. Indeed, the hadronic mass scale $(\Lambda_{\rm QCD})$ 
determining the mass of the proton, the neutron and the other strongly 
interacting particles is (via the Renormalization Group) predicted to 
be exponentially related to the string mass $m_s$. Roughly
\begin{equation}
m_p \sim m_s \, \exp (-b / \alpha_U) \label{eq8}
\end{equation}
where $b$ is a (known) number of order unity. Combining (\ref{eq8}) 
with (\ref{eq6}) leads to
\begin{equation}
\alpha_g (m_p) = \frac{G \, m_p^2}{\hbar c} \sim \alpha_U \, e^{-2b / 
\alpha_U} \, , \label{eq9}
\end{equation}
where $\alpha_U$ is the common value of the gauge coupling constants at 
unification. 

Finally, let me mention that String Theory (and other attempts at 
quantizing gravity, and/or unifying it with the other interactions) makes 
other generic predictions that might be testable in Cavendish-type 
experiments. Indeed, a generic prediction of such theories is that 
there are more gravitational-strength interactions than the usual 
(tensor) one described by Einstein's general relativity. In particular, 
the usual tensor gravitational field $g_{\mu \nu} (x)$ is typically 
accompanied by one or several scalar fields $\varphi (x)$. As many 
high-precision tests of relativistic gravity have put stringent limits 
on any long-range scalar gravitational fields (see, e.g., 
\cite{damour}), there are two possibilities (assuming that such scalar 
partners of $g_{\mu \nu}$ do exist in Nature): 

(i) the scalar 
gravitational field $\varphi (x)$ is (like $g_{\mu \nu}$) long-ranged, 
but its strength has been reduced much below the usual gravitational 
strength $G$ by some mechanism. [A natural cosmological mechanism for 
the reduction of any scalar coupling strength has been discussed in 
Refs.~\cite{dn}, \cite{dp}.]; 

(ii) the initially long-ranged field 
$\varphi (x)$, has acquired a mass-term $m_{\varphi}$, i.e. it has 
become short-ranged (decreasing with distance like $e^{-m_{\varphi} r} 
/ r$), but its strength is still comparable to (or larger than) $G$ 
\cite{tv}, \cite{ia}. 

In the first case, the best hope of detecting such a 
deviation from standard gravity is to perform ultra-high-precision 
tests of the equivalence principle \cite{dp}. In the second case, 
deviations from standard (Newtonian) gravity might appear in 
short-distance Cavendish-type experiments \cite{tv}, \cite{ia}.
 Indeed, it is 
possible (but by no means certain) that the mass (and therefore the 
range $m_{\varphi}^{-1}$) of such gravitational-strength fields be 
related to the supersymmetry breaking scale $m_{\rm SUSY}$ by a 
relation of the type $m_{\varphi} \sim G^{1/2} \, m_{\rm SUSY}^2 \sim 
10^{-3} \, {\rm e V} \, (m_{\rm SUSY} / {\rm Te} V)^2$. Therefore, if $m_{\rm 
SUSY} \sim 1 {\rm Te} V$, the observable strength of gravity would 
 increase  by a factor of order unity
at distances below $m_{\varphi}^{-1} \sim 1 \, {\rm mm}$ 
\cite{ia}, \cite{ferrara}. 
 More recently, another line of thought has  suggested that gravity could be 
even more drastically modified below some distance 
$r_0$ \cite{dimopoulos}. In principle, Cavendish-type experiments 
performed for separations smaller than $r_0$ might see a change 
of the $1/r^2$ law: the exponent 2 being replaced by an exponent larger 
than or equal to 4 \cite{dimopoulos} ! [Note, however, that in these 
models $r_0$ is already constrained to be smaller than $\sim 1 \mu {\rm m}$.] 
I wish also to mention a general argument of Weinberg \cite{weinberg} 
suggesting the existence of a new gravitational-related interaction 
with range {\it larger} than 0.1~mm \cite{weinberg}. [The recent 
announcement of the measurement of a non zero cosmological constant 
goes in the direction of confirming the importance of such a 
submillimeter scale.]

In conclusion, I hope to have shown that $G$-measurements and 
Cavendish-type experiments have now reached a new significance as 
possible windows on the physics of unification between gravity and the 
other interactions.


\begin{references}
\bibitem{cavendish}  H. Cavendish, Phil. Trans. Roy. Soc. {\bf 83} 
(1798) 470.
\bibitem{cwfe}  C.W.F. Everitt, {\it Theoretical significance of 
present-day gravity experiments}, in {\it Proceedings of the First 
Marcel Grossmann Meeting on General Relativity}, ed. R.~Ruffini 
(North-Holland, Amsterdam, 1977), pp.~545-615.
\bibitem{caso}  C. Caso et al., {\it Review of Particle Physics}, 
The European Physical Journal {\bf C3} (1998) 1.
\bibitem{polonsky}  See, e.g., N.~Polonsky, hep-ph/9411378; 
P.~Langacker, hep-ph/9412361.
\bibitem{landau}  L. Landau, in {\it Niels Bohr and the Development of 
Physics}, ed. W.~Pauli (McGraw-Hill, New York, 1955).
\bibitem{thooft} G. 't~Hooft, Nucl. Phys. B {\bf 315} (1989) 517.
\bibitem{corfu}  T. Damour, {\it Gravitation, Experiment and 
Cosmology}, gr-qc/9606079, to appear in the {\it Proceedings of the 5th 
Hellenic School of Elementary Particle Physics} (Corfu, September 
1995).
\bibitem{kaplu}  V.S. Kaplunovsky, Nucl. Phys. B {\bf 307} (1988) 145. 
[Erratum B {\bf 382} (1992) 436.]
\bibitem{dienes}  K.R. Dienes, Phys. Rep. {\bf 287} (1997) 447-528.
\bibitem{damour}  T. Damour, {\it Gravitation and Experiment}, in {\it 
Critical Problems in Physics}, ed. V.L.~Fitch et al. (Princeton Univ. 
Press, Princeton, 1997) pp.~147-166; gr-qc/9711061.
\bibitem{dn}  T. Damour and K. Nordtvedt, Phys. Rev. Lett. {\bf 70} 
(1993) 2217; Phys. Rev. D {\bf 48} (1993) 3436.
\bibitem{dp}  T. Damour and A.M. Polyakov, Nucl. Phys. B {\bf 423} 
(1994) 532; Gen. Rel. Grav. {\bf 26} (1994) 1171.
\bibitem{tv}  T.R. Taylor and G. Veneziano, Phys. Lett. B {\bf 213} 
(1988) 450.
\bibitem{ia}  I. Antoniadis, Phys. Lett. B {\bf 246} (1990)
377; I.~Antoniadis, S.~Dimopoulos and G.~Dvali, hep-ph/9710204.
\bibitem{ferrara}  S. Ferrara, C. Kounnas and F. Zwirner, Nucl. Phys. B 
{\bf 429} (1994) 589.
\bibitem{dimopoulos} N.~Arkani-Hamed, S.~Dimopoulos and G.~Dvali, 
hep-ph/9803315 and 9807344.
\bibitem{weinberg}  S. Weinberg, Rev. Mod. Phys. {\bf 61} (1989) 1-23.
\end{references}
\end{document}